\documentclass[12pt]{article} 
\usepackage{epsfig} 
\usepackage{amsfonts} 
\usepackage{amssymb}
\usepackage{amsmath}  
\usepackage{amsbsy}  
\usepackage{wasysym}
\usepackage{cite}
\usepackage{multirow} 
\input axodraw.sty
\catcode`\@=11 
\@addtoreset{equation}{section} 
\def\theequation{\thesection.\arabic{equation}} 
\@twosidetrue 
\catcode`\@=11  
\def\section{\@startsection{section}{1}{\z@}{3.5ex plus 1ex minus 
.2ex}{2.3ex plus .2ex}{\large\bf}} 
\def\thesection{\arabic{section}} 
\def\thesubsection{\arabic{section}.\arabic{subsection}} 
\def\thesubsubsection{\arabic{section}.\arabic{subsection}.\arabic{subsubsection}} 
\def\appendix{\setcounter{section}{0} 
\def\thesection{\Alph{section}} 
\def\theequation{\Alph{section}.\arabic{equation}} 
\def\thesubsection{\Alph{section}.\arabic{subsection}} 
\def\thesubsubsection{\Alph{section}.\arabic{subsection}.\arabic{subsubsection}} 
\def\section{\@startsection{section}{1}{\z@}{3.5ex plus 1ex minus 
   .2ex}{2.3ex plus .2ex}{\large\bf}} } 
\newcount\minutes 
\newcount\scratch 
\def\timestamp{%
\scratch=\time 
\divide\scratch by 60 
\edef\hours{\the\scratch} 
\multiply\scratch by 60 
\minutes=\time 
\advance\minutes by -\scratch 
---$\,$\hours:\null 
\ifnum\minutes< 10 0\fi 
\the\minutes} 

\def\sla#1{\ifmmode%
\setbox0=\hbox{$#1$}%
\setbox1=\hbox to\wd0{\hss$/$\hss}\else%
\setbox0=\hbox{#1}%
\setbox1=\hbox to\wd0{\hss/\hss}\fi%
#1\hskip-\wd0\box1 }

\def\wza{{$W^\pm Z\gamma$}}
\def\wpza{{$W^+ Z\gamma$}}
\def\wmza{{$W^- Z\gamma$}}
\def\vbfnlo{{\tt VBFNLO}}

\def\beq{\begin{equation}} 
\def\eeq{\end{equation}} 
 
\def\beqn{\begin{eqnarray}} 
\def\eeqn{\end{eqnarray}}

\def\({\left(} 
\def\){\right)} 
\def\as{\ifmmode \alpha_s \else $\alpha_s$ \fi}

\newcommand{\ie}{\textsl{i.e.}}

\begin{document} 
\begin{titlepage} 
\nopagebreak 
{\flushright{ 
        \begin{minipage}{5cm}
         FTUV--10--1109  \\            
         KA--TP--33--2010  \\            
         SFB/CPP-10-96 \\ 
         IFUM-967-FT \\
        \end{minipage} } 

} 
\vfill 
\begin{center} 
{\LARGE \bf 
 \baselineskip 0.5cm 
NLO QCD corrections to $W^\pm Z\gamma$ production with leptonic decays} 
\vskip 0.5cm  
{\large   
G. Bozzi$^1$, F. Campanario$^2$, M. Rauch$^2$, H. Rzehak$^2$ and D. Zeppenfeld$^2$
}   
\vskip .2cm
{\it $^1$ Dipartimento di Fisica, Universita' di Milano and INFN, Sezione di
Milano, \\
     20133 Milan, Italy}\\
{\it $^2$ Institut f\"ur Theoretische Physik, Karlsruhe Institute of Technology,\\
     Universit\"at Karlsruhe, 76128 Karlsruhe, Germany}\\
\vskip 1.3cm     
\end{center} 
\nopagebreak 
\begin{abstract}
We present a computation of the ${\cal O}(\alpha_s)$ QCD corrections to
\wza\ production at the Large Hadron Collider. The photon is considered
as real, and we include full leptonic decays for the $W$ and $Z$ bosons.
Based on the structure of the VBFNLO program package, we obtain
numerical results through a fully flexible Monte Carlo program, which
allows to implement general cuts and distributions of the final-state
particles.
The NLO QCD corrections are sizable and strongly exceed the theory
error obtained by a scale variation of the leading-order result. Also,
the shapes of relevant observables are significantly altered.
\end{abstract} 
\vfill 
\hfill 
\vfill 

\end{titlepage} 
\newpage               
%
%
\section{Introduction}
\label{sec:intro}

At the LHC a new energy frontier is reached, which allows for new
searches of unknown particles and further tests of the Standard
Model. From the theory side this also requires very precise predictions
of cross sections and distributions, exceeding the precision of a
leading-order approximation. We present a calculation of \wza\
production 
including QCD corrections and the leptonic decays of the $W$- and the
$Z$-boson with off-shell effects,  
\begin{align}
pp, p\bar{p} &\rightarrow  W^\pm Z \gamma +
X \rightarrow \ell_1^\pm
\stackrel{\text{\tiny(}-\text{\tiny)}}{\nu_1} \ell_2^+ \ell_2^- \gamma +
X\,.
\label{processes}
\end{align}
This process with leptons, photons and missing energy in the final state
provides a background to new physics searches (see, for example,
Ref.~\cite{Campbell:2006wx}). Also, this process
offers the possibility to study the quartic vector-boson couplings $W W Z
\gamma$ and $W W \gamma \gamma$ (see right diagram in Fig.~\ref{fig:1})
\cite{quartic} 
and test the 
Standard Model. It is one of the missing pieces for a full knowledge of
triple vector boson production  at next-to leading order (NLO)
QCD. There has been a strong effort for the calculation of these
processes. The processes with only massive vector bosons in the final
state have been completely calculated \cite{Lazopoulos:2007ix,
  Hankele:2007sb, Campanario:2008yg, Binoth:2008kt}. Rather recently,
also the NLO QCD calculation of the processes $W W \gamma$ and $Z Z \gamma$ have
been completed \cite{Bozzi:2009ig} and first results of the NLO QCD
computation of the process $W
\gamma \gamma$ have been presented \cite{Baur:2010zf}. Vector-boson pair
production accompanied with one jet has also been studied including QCD
corrections for $W W j$, $W \gamma j$, $W Z j$ and $Z Z j$
\cite{Campbell:2007ev, Dittmaier:2009un, Campanario:2009um, Campanario:2010hp, Binoth:2009wk}.

%
%
%
\section{Calculational Details}
\label{sec:calc}
\begin{figure}
\begin{center}
\includegraphics[scale=0.5]{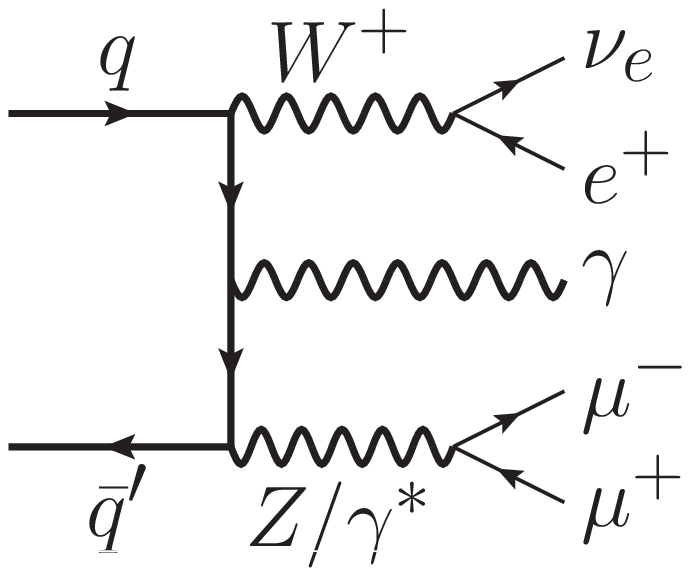}\hfill
\includegraphics[scale=0.5]{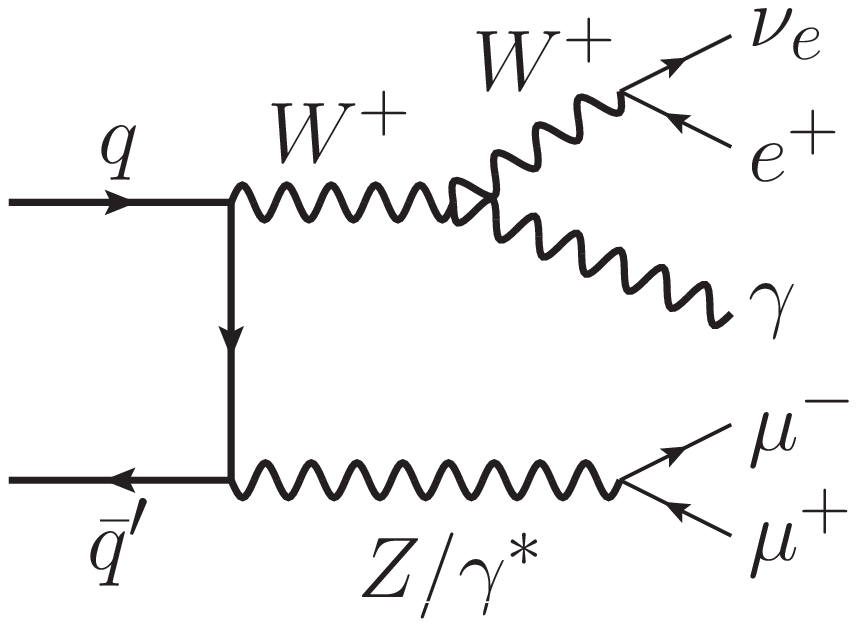}\hfill
\includegraphics[scale=0.5]{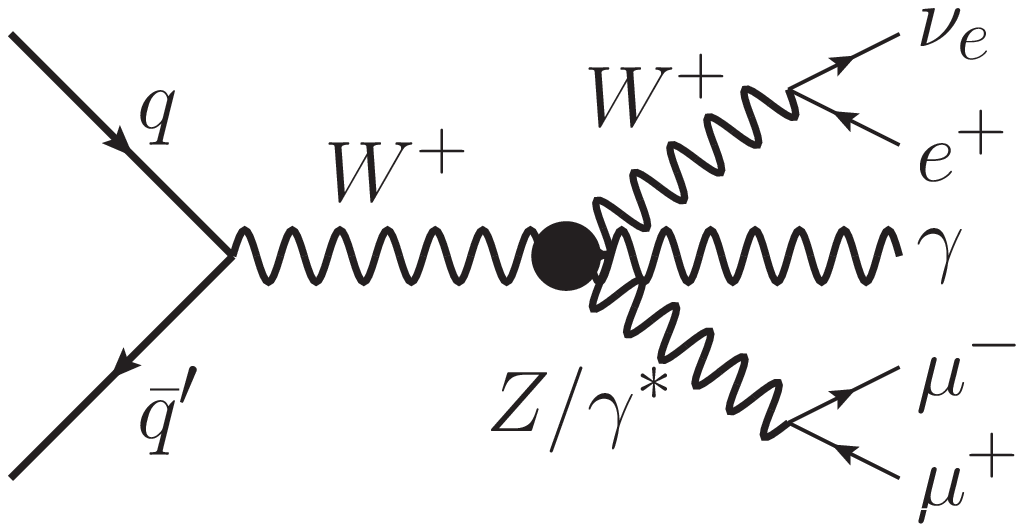}
\end{center}
\caption{Examples of the three topologies of Feynman diagrams
  contributing to the process $pp\to$\wza + X \, at tree-level. In the
  right-hand diagram, the quartic coupling is marked with a dot.}
\label{fig:1}
\end{figure}

We calculate all contributions to the processes (\ref{processes}) up to
order $\as\alpha^5$ in the limit where all fermions are massless. At
leading order, 71 distinct diagrams appear, which we group as three
different topologies, according to the number of vector bosons attached
to the quark line. An example of each class is depicted in
Fig.\ref{fig:1}.
For the last two topologies we also include the cases where vector bosons
are radiated off the lepton lines.

To speed up the calculation, invariant subparts, which appear multiple
times in different Feynman graphs, are computed only once per
phase-space point and independently of the rest of the cross-section.
Hereby, we use the procedure of {\it leptonic tensors}, as first
described in Ref.~\cite{Jager:2006zc}. This greatly reduces the
computation time needed.
For the computation of the matrix elements, we use the helicity method
introduced in Ref.~\cite{Hagiwara:1985yu}.
Furthermore, by charge conservation, a $W$ boson
must always couple to the quark line. Hence, we only need to compute the
left-handed chirality part.

At NLO QCD, virtual and real emission diagrams contribute to the cross
section.  Both 
contain infrared divergences, which must cancel in the sum according to
the Kinoshita-Lee-Nauenberg (KLN) theorem~\cite{Kinoshita:1962ur}. To
handle this cancellation in a numerically stable way, we use the
Catani-Seymour dipole subtraction method \cite{Catani:1996vz}.
Initial-state collinear singularities are partly factorized into the
parton-distribution functions. This leads to additional so-called
``finite collinear terms''.

The NLO real corrections are given by diagrams where an additional gluon
is attached to the quark line which is possible in two different ways.
Either this gluon is a final-state particle and considered as
radiated off the quark line, or an initial-state
gluon which splits into a $q\bar{q}$ pair and we have an emission of a
quark. With 194 different Feynman diagrams, the use of leptonic tensors proves
to be an advantage in this case.

The presence of isolated on-shell photons requires extra care in the
case of real emissions. An additional singularity arises from photon
emission collinear to a massless quark. Requiring a simple separation
cut between photon and jet is not allowed, since the cancellation of the
gluonic infrared divergences between virtual and real emission processes
would be spoiled. The phase space for soft-gluon emission
would be reduced while leaving untouched the virtual counterpart,
leading to the non-cancellation. In principle, this problem
can be solved by adding processes with quark fragmentation to photons.
Here, we use a simpler approach by requiring a specially crafted cut
first suggested in Ref.~\cite{Frixione:1998jh}. We will discuss the
details of this cut in section~\ref{sec:res}.

Virtual NLO QCD corrections arise from the insertion of gluon lines into
the topologies of Fig.~\ref{fig:1} in every possible way. Therefore, the
left topology gives rise to loops with up to five external particles,
\ie\ pentagon diagrams. For the middle and right ones at most box and
triangle diagrams, respectively, appear.  Up to the box level, we compute
the loop integrals using Passarino-Veltman
reduction~\cite{Passarino:1978jh}; additionally we avoid the explicit
calculation of the inverse of the Gram matrix. Instead, we solve a system
of linear equations, which is numerically more stable close to the
singular points. For the pentagon contributions, we apply the method of
Ref.~\cite{Denner:2002ii}. This circumvents the appearance of small Gram
determinants in planar configurations of the external momenta
altogether. The complete virtual corrections 
\begin{align}
M_V = \widetilde{M}_V + \ \frac{\alpha_S}{4 \pi} \ C_F \ \left( \frac{4
      \pi \mu^2}{Q^2} \right)^\epsilon  \ \Gamma{(1 + \epsilon)} \ \left[
    -\frac{2}{\epsilon^2} - \frac{3}{\epsilon} - 8 + \frac{4 \pi^2}{3}
  \right] \ M_B, \label{eq:MV}
\end{align}
can be separated into a part which is proportional to the Born matrix
element, $M_B$, and a remainder, $\widetilde{M}_V$. Here, $Q$ denotes the
partonic center-of-mass energy, which corresponds to the invariant mass
of the $WZ\gamma$ final state. For diagrams with only a single $W$ 
attached to the quark line, $\widetilde{M}_V$ vanishes, \ie\ the virtual
corrections completely factorize to the Born amplitude. 

For this factorization formula to hold, the transversality property of
the photon $\epsilon(p_\gamma) \cdot p_\gamma = 0$ must be
used~\cite{Bozzi:2009ig}. Then we can reuse the general results for the
finite remainder already obtained in
Refs.~\cite{Campanario:2008yg,Bozzi:2009ig}, adapting the attached
vector bosons to our case.  Another optimization can be performed by
shifting parts from the pentagon diagrams into box
contributions~\cite{Jager:2006zc,Hankele:2007sb,Campanario:2008yg,Bozzi:2009ig}.
To do this, we split the polarization vector of the vector bosons into a
part proportional to the four-momentum of the boson $q_V$ and a
remainder $\tilde{\epsilon}_V^\mu$, which is chosen such that
\begin{align}\label{eq:pentshift}
\tilde{\epsilon}_{V} \cdot (q_{W} + q_{Z}) = 0 \, .
\end{align}
This part leads to a reduction in the size of the pentagon
contributions, which we can therefore compute with lower statistics.
Contracting the pentagons with $q_V^\mu$, they can be expressed as a
difference of two box diagrams which are numerically faster to compute.
Hence, we can gain speed while keeping the total error the same.
This shift also serves as a consistency check between the box and
pentagon routines, as the total result must stay unchanged.

To verify the correctness of our calculation, we have performed several
checks. First, we have compared all tree-level amplitudes against matrix
elements generated by {\tt MadGraph}\cite{Stelzer:1994ta} and find an
agreement of at least 14 digits, which is at the level of the machine
precision. Additionally, we have also compared the integrated cross
section for both $pp\rightarrow W^\pm Z\gamma$ and $pp\rightarrow
W^\pm Z\gamma j$ against \texttt{MadEvent} and
\texttt{Sherpa}~\cite{Sherpa:09}. We find an agreement at the per mill
level, which is compatible with the integration error. Furthermore, we
have checked the implementation of the Catani-Seymour subtraction
scheme. We have verified that for the real emission part the ratio
between the differential real-emission cross section and the dipole
approaches $-1$ once we go to the soft or collinear limit. Also, we have
checked that finite contributions left after the cancellation of the
infrared divergence can be shifted between the virtual and real parts
without affecting the total result.

For the phase space implementation, the $2\rightarrow 5$ phase space was
split into three different parts. They correspond to three
topologies which we have identified to give the largest contributions,
which are of comparable size each. The first one corresponds to a
$2\rightarrow2$ s-channel structure, where possibly a massless particle
(the final-state jet for the real-emission contribution) is radiated off
first.  The two outgoing particles are the massless photon and a
pseudo-particle with a mass of 450 GeV and a width of 100 GeV. The
latter one then decays into the two massive vector bosons, which in turn
decay into their respective fermion--anti-fermion pair, all via
$1\rightarrow2$ topologies. For all invariant masses of intermediate
particles we apply a Breit-Wigner mapping to the corresponding random
number, so that the peak is flattened out, integrating over the whole
available energy range.
The second and third structure correspond to $2\rightarrow2$ s-channel
production of the two massive vector bosons, where one undergoes a
two-body decay into
its fermions and the other one, via a $1\rightarrow3$ process, into the
fermions and the photon.
To avoid double-counting 
we compute for both fermion--anti-fermion pairs the difference of the
invariant mass of the photon and the fermion--anti-fermion pair and the
mass of the corresponding vector boson. If both differences are larger
than 30 GeV, the phase-space point is assigned to the first topology,
otherwise to that one where the difference is smaller.

\section{Results}
\label{sec:res}
We perform the numerical evaluation of our calculation with an NLO Monte
Carlo program based on the structure of the \vbfnlo\ program
package~\cite{Arnold:2008rz}. As input parameters 
in the electroweak sector we take the $W$ and $Z$ boson masses
and the Fermi constant. The weak mixing angle and the
electromagnetic coupling constant are computed from these using
tree-level relations:
\begin{align}
&m_W = 80.398 \ \mathrm{GeV} && \sin^2{(\theta_W)} = 0.22264 \nonumber\\ 
&m_Z = 91.1876 \ \mathrm{GeV} && \alpha^{-1} = 132.3407 \nonumber\\
&G_F = 1.16637 \cdot 10^{-5} \ \mathrm{GeV}^{-2}  
\; . &&  \label{eq:ew}
\end{align}

Top-quark effects are not considered and all other quarks are taken massless.
Effects from generation mixing are neglected, as we set the CKM matrix
to the identity matrix. As the central value for factorization and
renormalization scales we choose the invariant mass of the leptons and
the photon
\begin{align} \label{eq:WZAmass}
\mu_F = \mu_R = \mu_0 =  m_{WZ\gamma} \equiv
\sqrt{(p_{\ell_1} + p_{\nu_1} + p_{\ell_2} + 
  p_{\bar{\ell}_2} + p_{\gamma})^2}.
\end{align}
For the parton distribution functions, we choose CTEQ6L1 at LO and the
CTEQ6M set with $\alpha_S(m_Z)=0.1176$ at NLO~\cite{Pumplin:2002vw}.

We impose the following set of minimal cuts on the rapidity $y$
and the transverse momentum, $p_T$, of the final-state photon and charged
leptons
\begin{align}
p_{T_{\gamma}} &> 10 \ \mathrm{GeV} &
p_{T_{\ell}} &> 20 \ \mathrm{GeV} &
|y_{\gamma}| &< 2.5 &
|y_{\ell}| &< 2.5 \ .
\label{eq:cuts1}
\end{align}
These take into account typical requirements of the experimental detectors.
Furthermore, leptons, photon and jet need to be well separated in order
to avoid divergences from collinear photons and to be able to identify
them as separate objects in the detectors. Therefore, we impose
\begin{align}
R_{\ell\ell} &> 0.3  &
R_{\ell\gamma} &> 0.4  &
R_{j \ell} &> 0.4  &
R_{j \gamma} &> 0.4 &
m_{\ell\ell} &> 15 \ \mathrm{GeV} \ .
\label{eq:cuts2}
\end{align}
Here, a jet refers to a final-state quark or gluon in the NLO real
emission contribution with $p_{T_j} > 20$ GeV and $|y_{j}| < 4.5$.
The last cut in Eq.~(\ref{eq:cuts2}) eliminates the singularity from a
virtual photon $\gamma^* \rightarrow \ell^+ \ell^-$ by requiring that
the invariant mass of each pair of oppositely charged leptons is larger
than 15 GeV.
For treating the collinear singularity between the photon and a parton $i$,
we use the procedure of Ref.~\cite{Frixione:1998jh}. The event is
accepted only if
\begin{equation}
p_{T_i} \le p_{T_{\gamma}} \frac{1-\cos R_{\gamma i}}{1-\cos\delta_0} 
\text{\quad or\quad} R_{\gamma i} > \delta_0 \ ,
\label{eq:frixione}
\end{equation}
where $\delta_0$ is a fixed separation parameter which we set to $0.7$.
Eq.~(\ref{eq:frixione}) allows final-state partons arbitrarily close to the
photon axis as long as they are soft enough. Thereby, it retains the full
QCD pole, which cancels against the virtual part, but it does not
introduce divergences from the electroweak sector.

\begin{table}
\begin{center}
\begin{tabular}{|l|lll|lll|}
\hline
\multirow{2}{*}{LHC}
 & \multicolumn{3}{|c|}{$\sqrt{s}=$7 TeV}
 & \multicolumn{3}{|c|}{$\sqrt{s}=$14 TeV} \\ 
 & \multicolumn{1}{|c}{LO[fb]} & \multicolumn{1}{c}{NLO [fb]} & \multicolumn{1}{c|}{K-factor} 
 & \multicolumn{1}{|c}{LO[fb]} & \multicolumn{1}{c}{NLO [fb]} & \multicolumn{1}{c|}{K-factor} \\ \hline
$4 \, \sigma("W^+Z\gamma" \to e^+ \nu_e \mu^+ \mu^- \gamma)$ &&&&&& \\
$~~p_{T_{\gamma(\ell)}}>10(20)$ GeV 
&\multirow{1}{*}{~~0.650~~} &\multirow{1}{*}{~~1.075~~} &\multirow{1}{*}{~~1.65~~}
&\multirow{1}{*}{~~1.324~~} &\multirow{1}{*}{~~2.441~~} &\multirow{1}{*}{~~1.84~~}\\ 
$~~p_{T_{\gamma(\ell)}}>20(20)$ GeV 
&\multirow{1}{*}{~~0.278~~} &\multirow{1}{*}{~~0.482~~} &\multirow{1}{*}{~~1.74~~}
&\multirow{1}{*}{~~0.587~~} &\multirow{1}{*}{~~1.169~~} &\multirow{1}{*}{~~1.99~~}\\ 
\hline
$4 \,\sigma("W^-Z\gamma" \to e^- \bar{\nu}_e \mu^+ \mu^- \gamma)$ &&&&&& \\
$~~p_{T_{\gamma(\ell)}}>10(20)$ GeV 
&\multirow{1}{*}{~~0.352~~} &\multirow{1}{*}{~~0.621~~} &\multirow{1}{*}{~~1.76~~}
&\multirow{1}{*}{~~0.886~~} &\multirow{1}{*}{~~1.717~~} &\multirow{1}{*}{~~1.94~~}\\ 
$~~p_{T_{\gamma(\ell)}}>20(20)$ GeV 
&\multirow{1}{*}{~~0.146~~} &\multirow{1}{*}{~~0.275~~} &\multirow{1}{*}{~~1.88~~}
&\multirow{1}{*}{~~0.381~~} &\multirow{1}{*}{~~0.813~~} &\multirow{1}{*}{~~2.13~~}\\ 
\hline
 \end{tabular}
\caption[]{Total cross sections at the LHC with center-of-mass energies
of 7 and 14 TeV for $pp \to $ \wza$+X$ with leptonic decays, at LO and
NLO, and for two sets of cuts. Relative statistical errors of the Monte
Carlo are at the per mill level. The factor 4 accounts for all
combinations of final-state first- and second-generation leptons.}
\label{LHCnumbers}
 \end{center}
 \end{table}
\begin{table}
\begin{center}
\begin{tabular}{|l|lll|}
\hline
Tevatron ($\sqrt{s} = 1.96$ TeV)
 & \multicolumn{1}{|c}{LO[ab]} & \multicolumn{1}{c}{NLO [ab]} & \multicolumn{1}{c|}{K-factor} \\ \hline
$4 \,\sigma("W^\pm Z\gamma" \to e^\pm \stackrel{\text{\tiny(}-\text{\tiny)}}{\nu}_e \mu^+ \mu^- \gamma)$ &&& \\
$~~p_{T_{\gamma(\ell)}}>10(10)$ GeV 
&\multirow{1}{*}{~~250.0~~} &\multirow{1}{*}{~~370.8~~} &\multirow{1}{*}{~~1.48~~}\\ 
$~~p_{T_{\gamma(\ell)}}>20(10)$ GeV 
&\multirow{1}{*}{~~107.4~~} &\multirow{1}{*}{~~160.1~~} &\multirow{1}{*}{~~1.49~~}\\ 
\hline
 \end{tabular}
\caption[]{Total cross sections at the Tevatron for $p\bar p \to $
  \wpza$+X$ or $p\bar p \to $ \wmza$+X$ including leptonic decays, at
  LO and NLO, and for two sets of cuts. Relative statistical errors of
  the Monte Carlo are below the per mill level. The factor 4 accounts
  for all combinations of final-state first- and second-generation
  leptons.}
\label{Tevnumbers}
 \end{center}
 \end{table}

In Tables~\ref{LHCnumbers} and~\ref{Tevnumbers}, we present results for
the integrated cross section of \wza{} production for the LHC with a
center-of mass energy of both $7$ and $14$ TeV and for the Tevatron with
its center-of-mass energy of $1.96$ TeV. Note, for the Tevatron, the
cross section for $W^+ Z \gamma$ and $W^- Z \gamma$ production is the same; 
the given number is the individual result of one of them.  Besides
the standard cut on $p_{T_\gamma}$ of 10 GeV we also show results for 20
GeV. For the Tevatron, we have reduced the cut on $p_{T_\ell}$ to 10 GeV
throughout.  The cross sections shown correspond to the production of
both electrons and muons for all leptons. Interference effects from
identical leptons in the final state are neglected, since their
contribution is small.

\begin{figure}[tbp]
\includegraphics[width=0.48\textwidth]{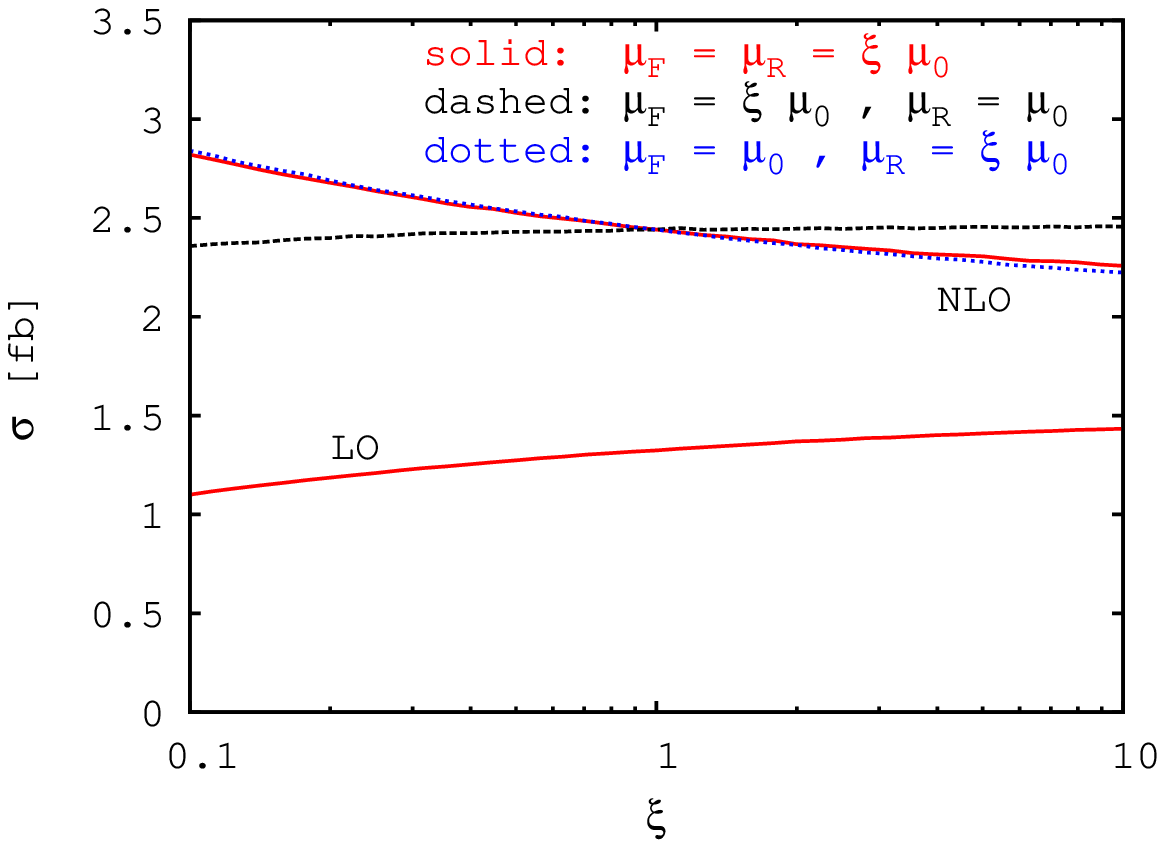}
\includegraphics[width=0.48\textwidth]{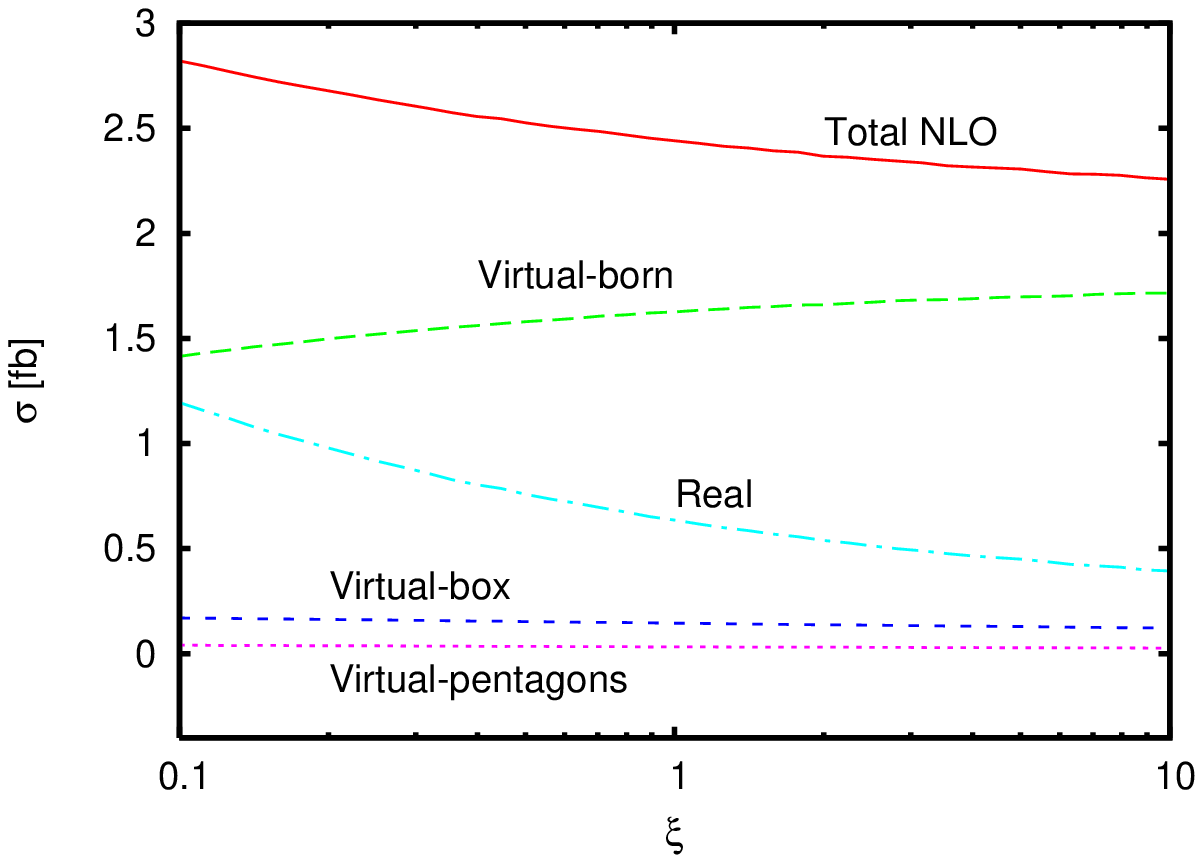}
\caption[]{\label{fig:scaleplus}
{\it Left:} {\sl Scale dependence of the total LHC cross section  at
  $\sqrt{s} = 14$~TeV for 
  $p p \to W^+Z\gamma+X \to \ell_1^+ \ell_2^+ \ell_2^- \gamma +\sla{p}_T+X$ at LO and
  NLO within the cuts of
Eqs.~(\ref{eq:cuts1}, \ref{eq:cuts2}, \ref{eq:frixione}).
  The factorization and renormalization scales are together or
  independently varied in the range from $0.1 \cdot \mu_0$ to $10 \cdot
  \mu_0$.} 
{\it Right:} {\sl Same as in the left panel but for the different NLO
  contributions at $\mu_F=\mu_R=\xi\mu_0$ with $\mu_0 = m_{W Z \gamma}$.}}
\end{figure}
\begin{figure}[tbp]
\includegraphics[width=0.48\textwidth]{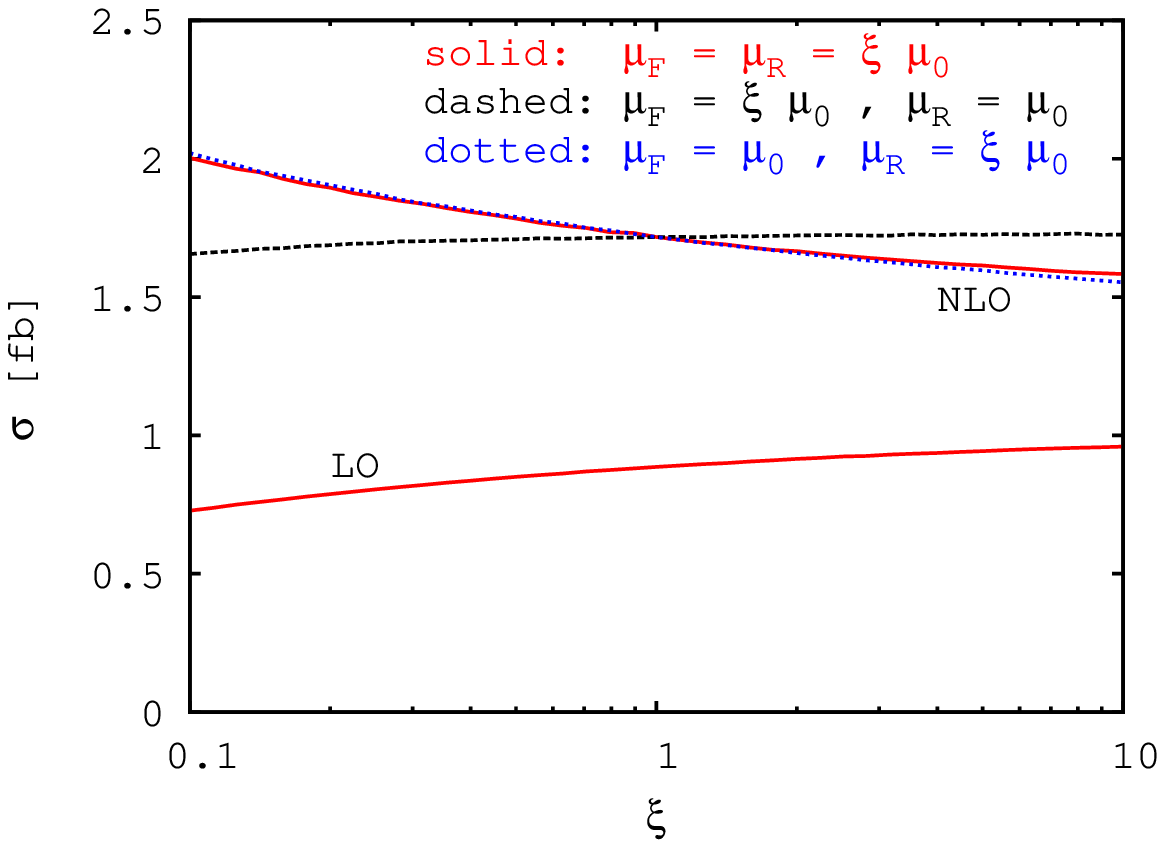}
\includegraphics[width=0.48\textwidth]{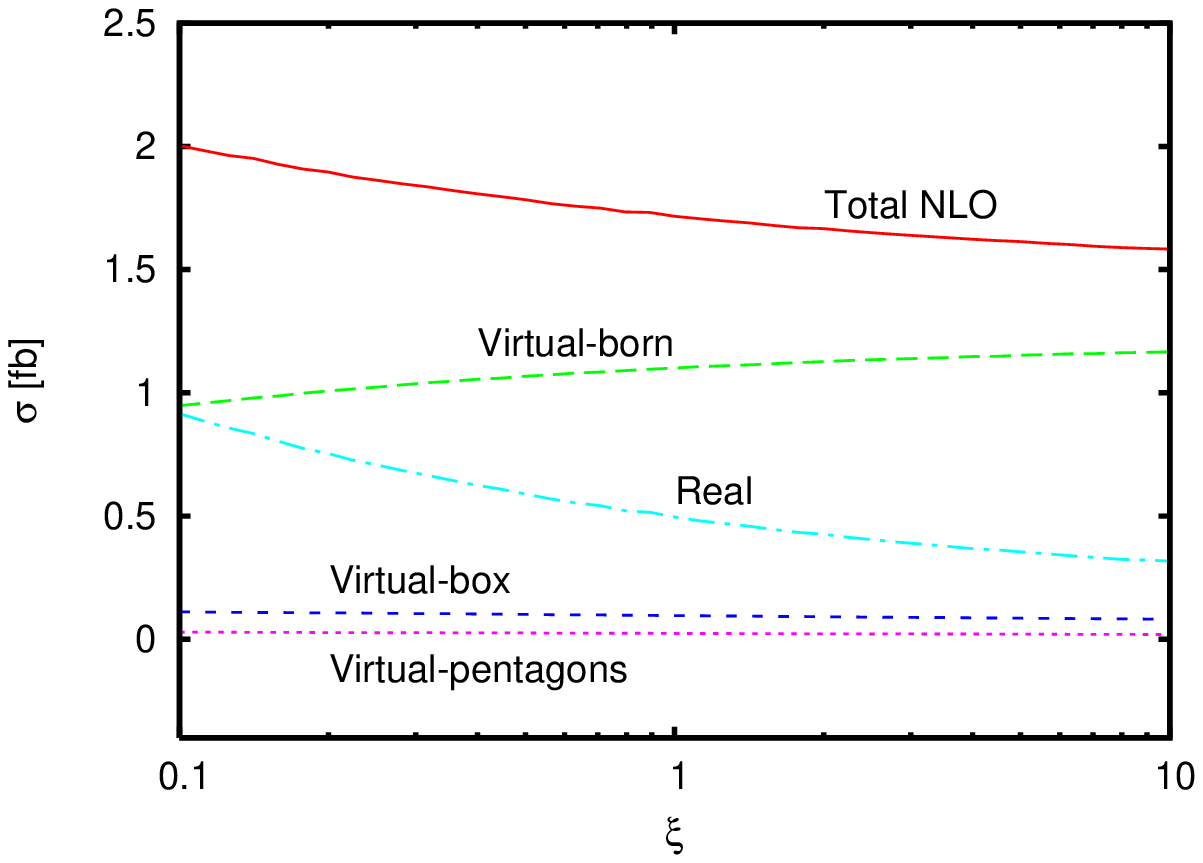}
\caption[]{\label{fig:scaleminus}
{\it Left:} {\sl Scale dependence of the total LHC cross section at
  $\sqrt{s} = 14$~TeV for 
  $p p \to W^-Z\gamma+X \to \ell^-_1\ell^+_2 \ell^-_2  \gamma+\sla{p}_T+X+X $ 
  at LO and NLO within the cuts of
Eqs.~(\ref{eq:cuts1}, \ref{eq:cuts2}, \ref{eq:frixione}).
  The factorization and renormalization scales are together or
  independently varied in the range from $0.1 \cdot \mu_0$ to $10 \cdot
  \mu_0$.} 
{\it Right:} {\sl Same as in the left panel but for the different NLO
  contributions at $\mu_F=\mu_R=\xi\mu_0$ with $\mu_0 = m_{W Z \gamma}$.}}
\end{figure}

From hereon, we will focus on the LHC with a center-of-mass energy of 14
TeV.
In Figs.~\ref{fig:scaleplus} and~\ref{fig:scaleminus} we show the
dependence of the cross section for \wpza{} and \wmza{} production,
respectively, when varying the renormalization and factorization scale
in the interval
\begin{equation}
\mu_F, \mu_R = \xi \cdot \mu_0 \quad (0.1 < \xi < 10)
\end{equation}
around the central scale $\mu_0$ given in Eq.~(\ref{eq:WZAmass}).
We see that the variation of the LO cross section with the scale
strongly underestimates the size of the NLO contributions. At the central
scale, we obtain a K-factor of $1.84$ for \wpza{} and $1.94$ for \wmza{}.
The dependence on the factorization scale slightly reduces 
when we move from a LO calculation to NLO as expected. On the
other hand, the dependence on the renormalization scale shows a large
variation. This is due to the fact that $\alpha_s$
enters the cross section only at NLO, where we observe the typical
leading renormalization scale dependence. 
When varying the factorization and the renormalization scale jointly by
a factor 2 around the central scale $\mu_0$, we see a change of 7.5\% at
LO and 6.7\% at NLO for \wpza{}. For \wmza{}, the numbers change only
slightly to 7.7\% and 7.3\% for LO and NLO, respectively.

On the right-hand side of Figs.~\ref{fig:scaleplus}
and~\ref{fig:scaleminus}, we show the combined factorization and
renormalization scale dependence of the NLO cross section split into the
individual contributions.  Almost the entire scale dependence is given
by the real-emission part, which contains the true real-emission cross
section, the dipole terms from the Catani-Seymour subtraction scheme and
the finite collinear terms. We obtain the bulk of the NLO contribution 
from the Born matrix element and the virtual corrections proportional to
it. This includes the terms from boxes and pentagons
factored out in Eq.~(\ref{eq:MV}). At the central scale, it is more than
twice as large as the real part. The finite virtual remainders due to
box and pentagon corrections,  which are shifted using
Eq.~(\ref{eq:pentshift}), only yield a small contribution. The
shape of their scale dependence is similar to the one of the total
cross section.

\begin{figure}[tbp]
\includegraphics[width=0.48\textwidth]{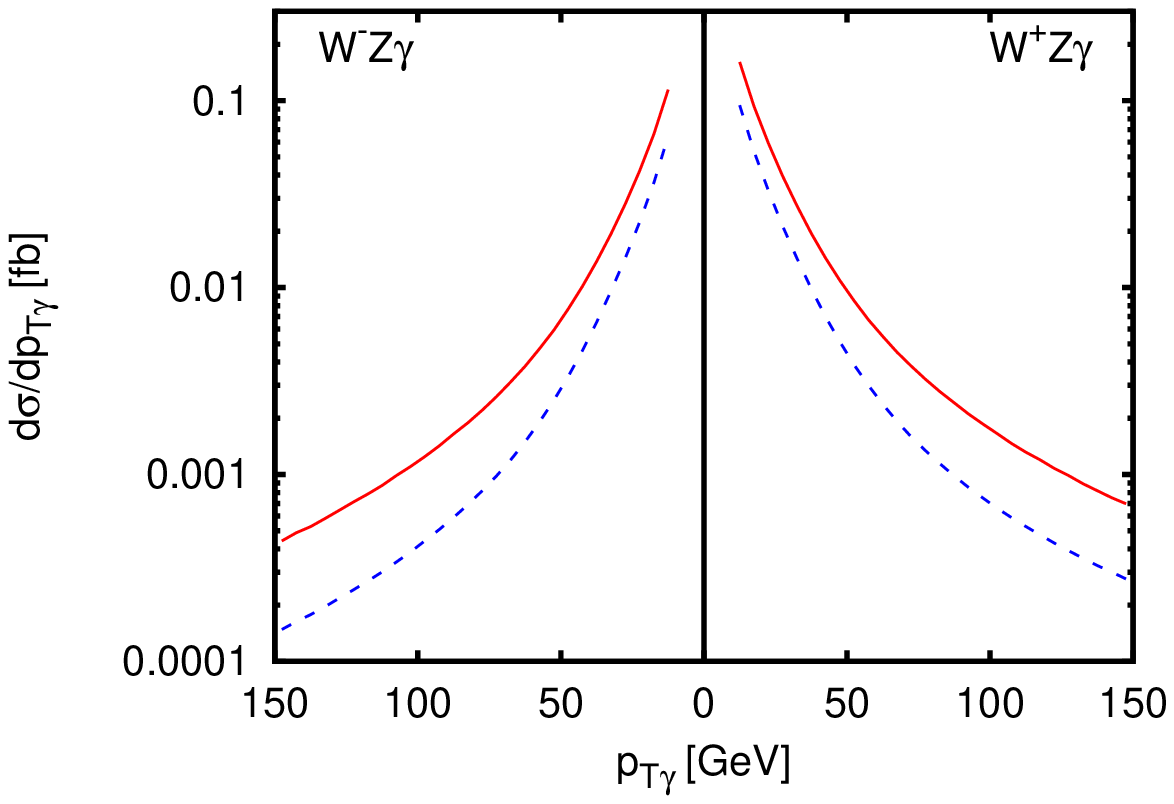}
\includegraphics[width=0.48\textwidth]{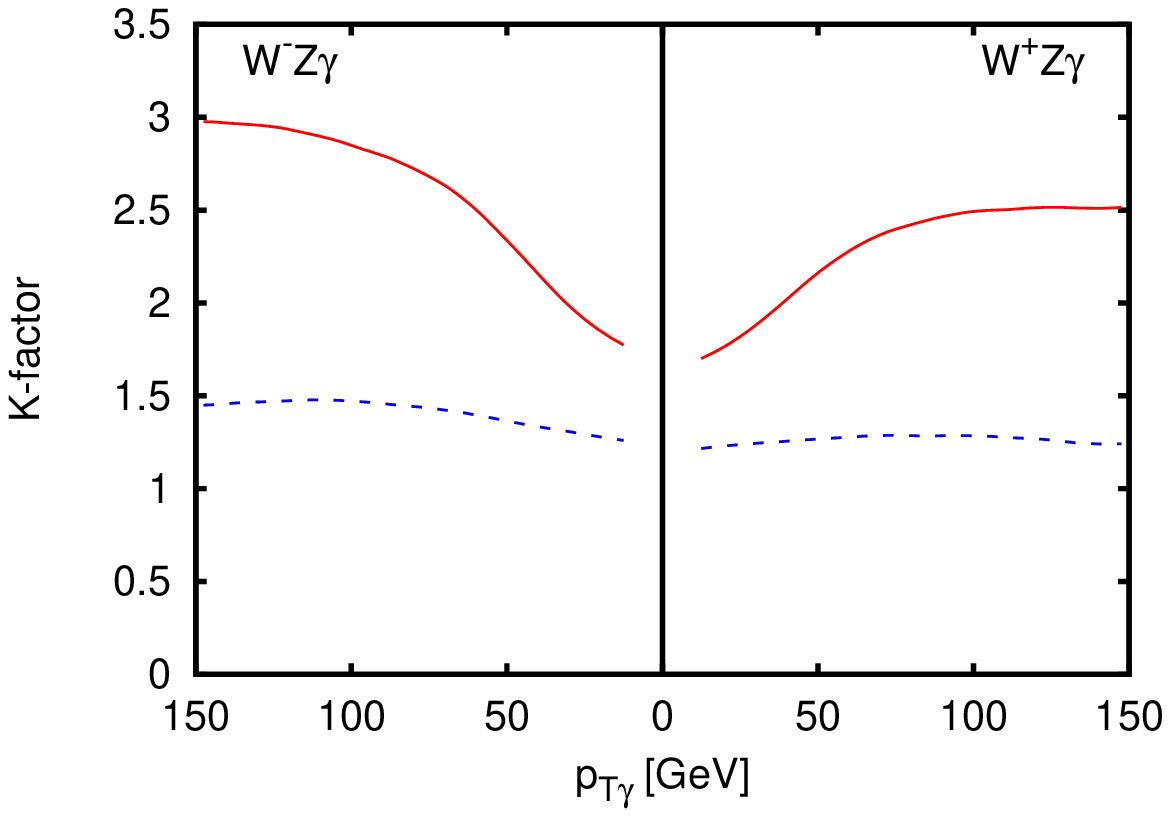}
\caption[]{\label{fig:ptgamma}
{\it Left:} {\sl Transverse-momentum distribution of the photon in
\wpza\, and \wmza\, production with leptonic decays of the W- and the
Z-boson at the LHC with $\sqrt{s} = 14$~TeV. LO (dashed blue line) and
NLO (solid red) results are shown for $\mu_F=\mu_R=\mu_0=m_{WZ\gamma}$
and the cuts of Eqs.~(\ref{eq:cuts1}, \ref{eq:cuts2}, \ref{eq:frixione}).} 
{\it Right:} {\sl K-factor for the transverse-momentum distribution of
the photon as defined in Eq.~(\ref{eq:kfactor}) without (solid red) and
including a jet veto of 50 GeV (dashed blue).}}
\end{figure}
\begin{figure}[tbp]
\includegraphics[width=0.48\textwidth]{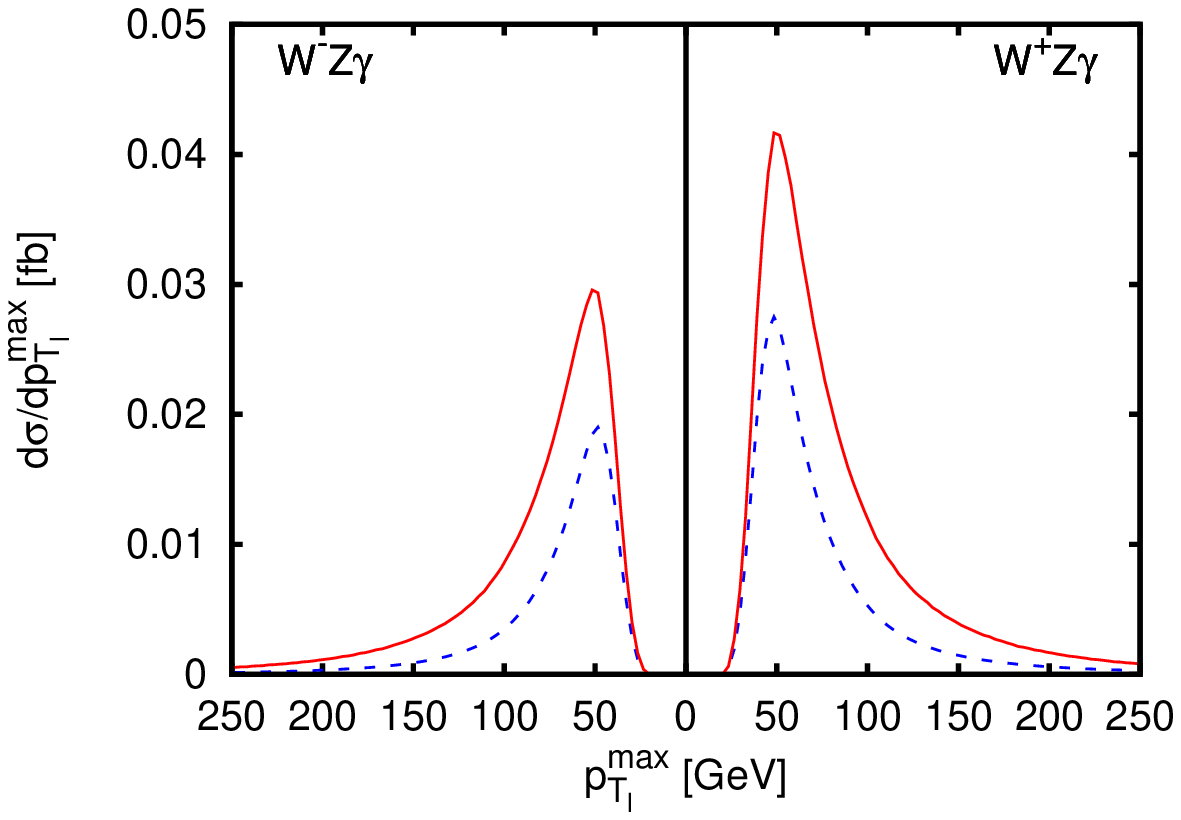}
\includegraphics[width=0.48\textwidth]{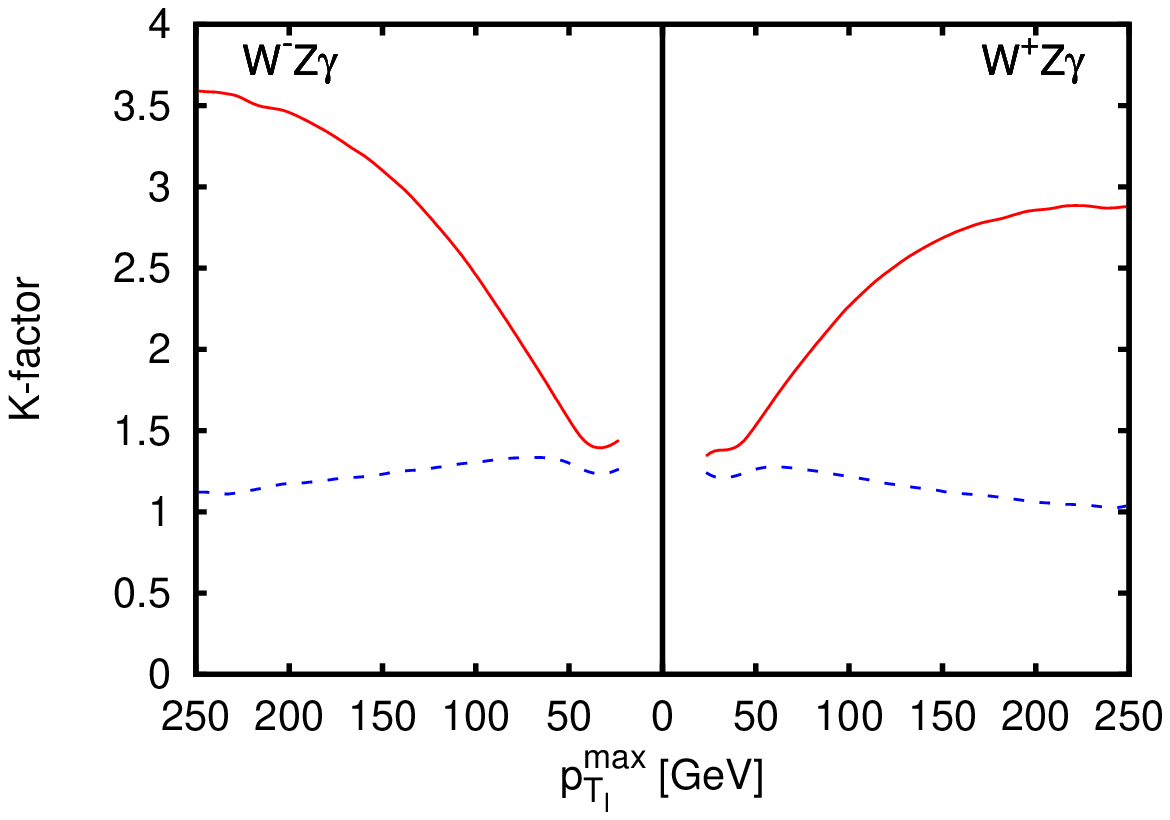}
\caption[]{\label{fig:ptl1}
{\it Left:} {\sl Transverse-momentum distribution of the lepton with
largest transverse momentum in \wpza\, and \wmza\, production  with
leptonic decays of the W- and the Z-boson at the LHC with $\sqrt{s} =
14$~TeV. LO (dashed blue line) and NLO (solid red) results are shown for
$\mu_F=\mu_R=\mu_0=m_{WZ\gamma}$ and the
cuts of Eqs.~(\ref{eq:cuts1}, \ref{eq:cuts2}, \ref{eq:frixione}).} 
{\it Right:} {\sl K-factor for the maximal-transverse-momentum
distribution of the lepton as defined in Eq.~(\ref{eq:kfactor}) without
(solid red) and including a jet veto of 50 GeV (dashed blue).}}
\end{figure}
\begin{figure}[tbp]
\includegraphics[width=0.48\textwidth]{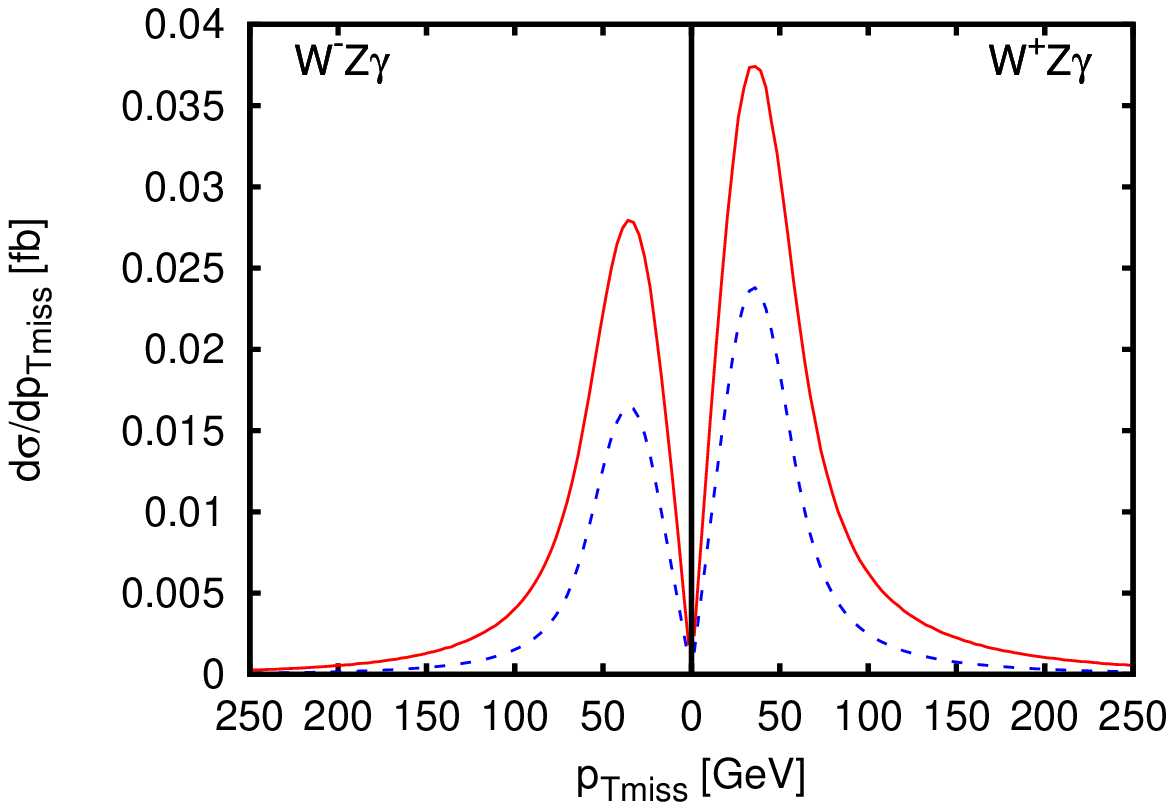}
\includegraphics[width=0.48\textwidth]{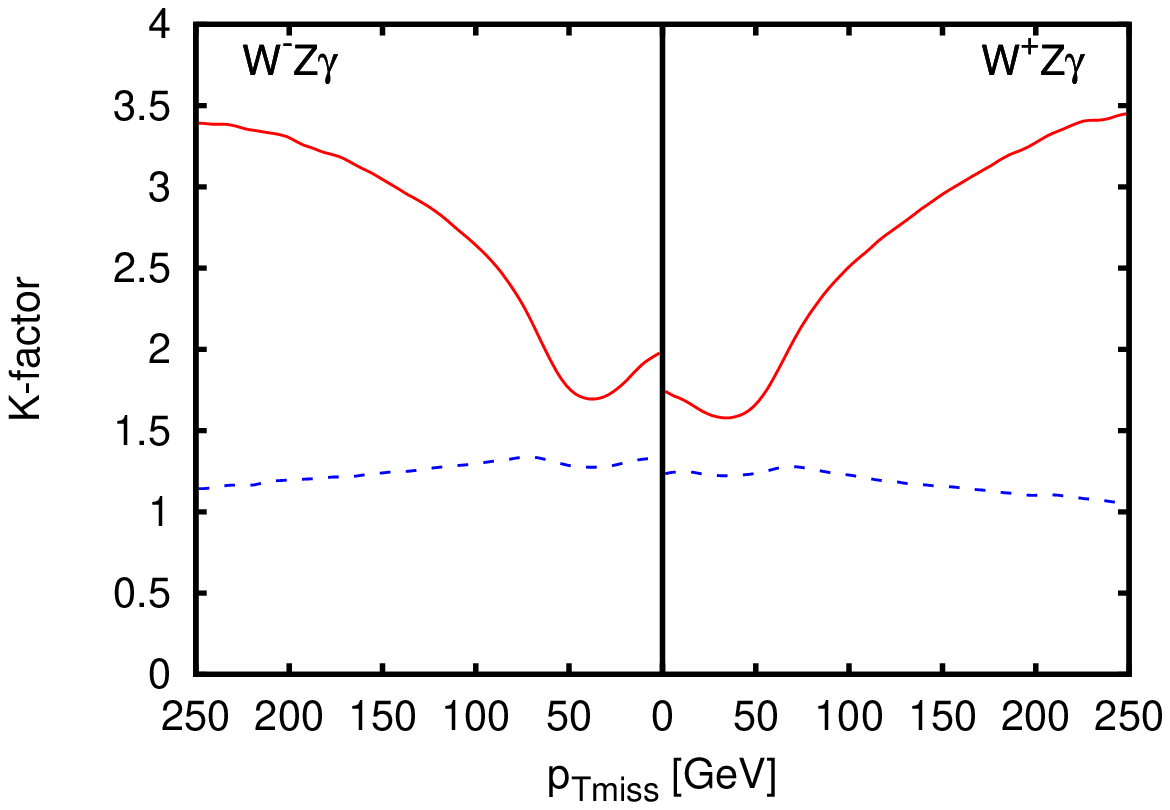}
\caption[]{\label{fig:ptmiss}
{\it Left:} {\sl Distribution of missing transverse momentum in \wpza\,
and \wmza\, production with leptonic decays of the W- and the Z-boson at
the LHC with $\sqrt{s} = 14$~TeV. LO (dashed blue line) and NLO (solid
red) results are shown for $\mu_F=\mu_R=\mu_0=m_{WZ\gamma}$
and the cuts of Eqs.~(\ref{eq:cuts1}, \ref{eq:cuts2}, \ref{eq:frixione}).} 
{\it Right:} {\sl K-factor for the missing transverse-momentum
distribution as defined in Eq.~(\ref{eq:kfactor}) without (solid red) and
including a jet veto of 50 GeV (dashed blue).}}
\end{figure}
\begin{figure}[tbp]
\includegraphics[width=0.48\textwidth]{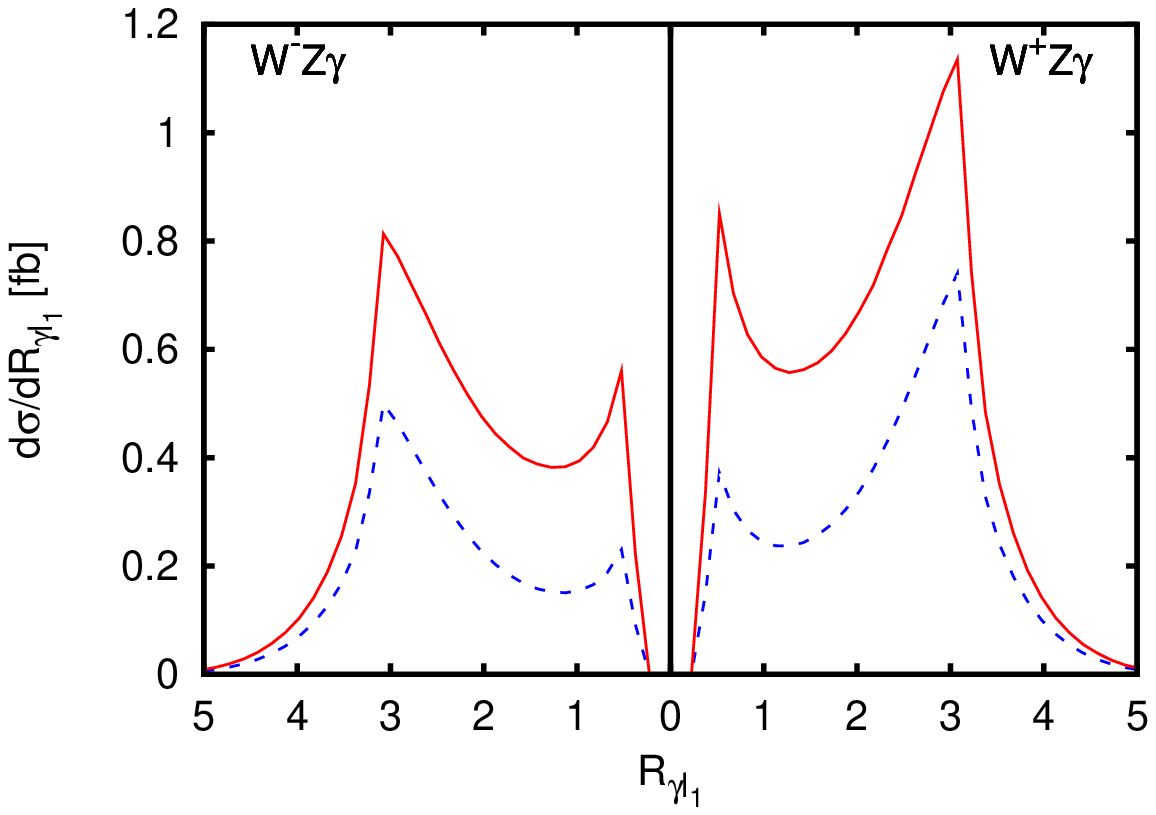}
\includegraphics[width=0.48\textwidth]{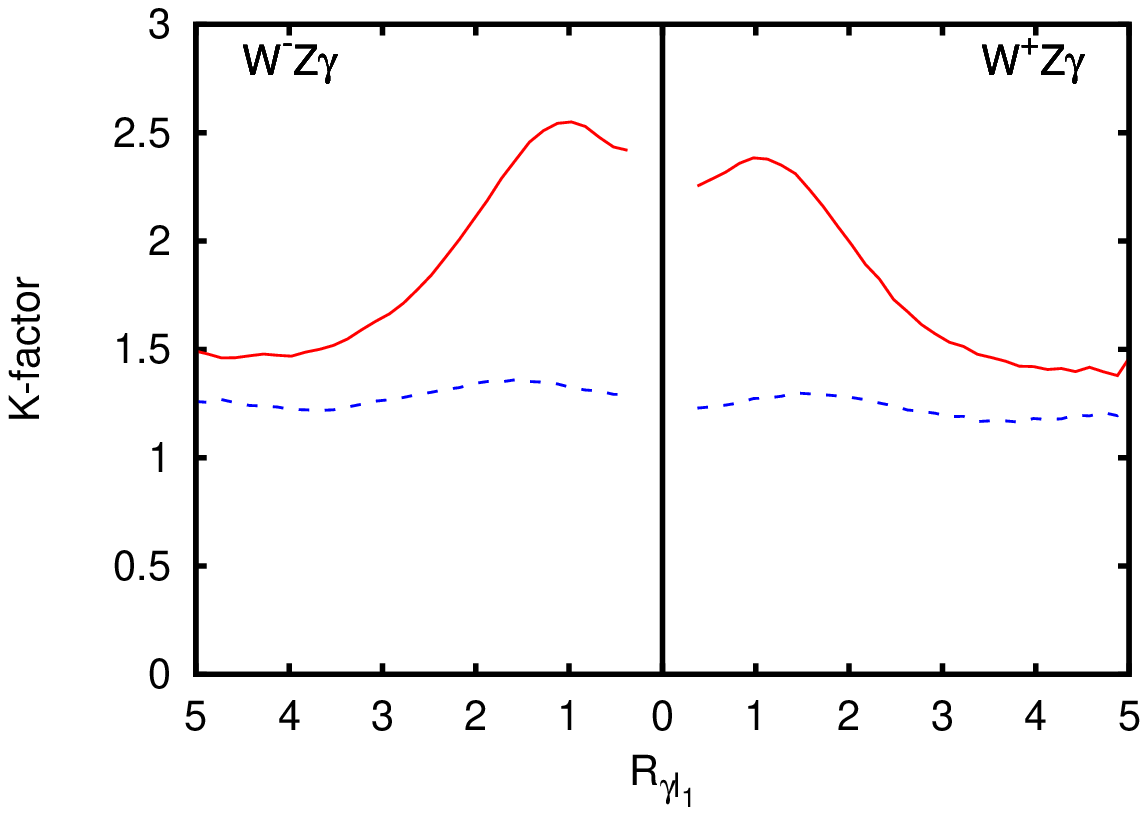}
\caption[]{\label{fig:RAl1}
{\it Left:} {\sl Separation of the photon and the hardest lepton in
\wpza\, and \wmza\, production with leptonic decays of the W- and the
Z-boson at the LHC with $\sqrt{s} = 14$~TeV. LO (dashed blue line) and NLO
(solid red) results are shown for $\mu_F=\mu_R=\mu_0=m_{WZ\gamma}$ and
the cuts of Eqs.~(\ref{eq:cuts1}, \ref{eq:cuts2}, \ref{eq:frixione}).} 
{\it Right:} {\sl K-factor for the distribution of the photon-lepton
separation as defined in Eq.~(\ref{eq:kfactor}) without (solid red) and
including a jet veto of 50 GeV (dashed blue).}}
\end{figure}

In Figs.~\ref{fig:ptgamma}, \ref{fig:ptl1} and~\ref{fig:ptmiss}, we show
the distribution of the transverse momentum of the photon and the
hardest lepton as well as of the missing transverse momentum originating
from the neutrino, respectively.
On the left-hand side of each figure, we depict the differential cross
section at LO and NLO both for \wpza{} and \wmza{}, and on the right-hand
side we plot the differential K-factor, defined in the following
way:
\begin{align}\label{eq:kfactor}
K = \frac{d \sigma^{NLO}/ dx}{d \sigma^{LO} /dx} \,\, .
\end{align}
We present results which do not include any cut on the additional jet,
as well as including a veto on jets with $p_{T_j} > 50$ GeV. As
previously, a jet is defined as a final-state parton with
$|y_j|<4.5$.

We see that for each of the distributions the K-factor for the
differential cross sections without jet veto is not constant,
but shows a strong dependence on the momentum scale. In all cases it is
close to the integrated one for small values of the transverse momenta
coinciding with the bulk of the cross section. For large transverse momenta
we observe much bigger K-factors, typically extending up to a value of
three. 
Once we include the additional jet veto, this strong dependence is
largely removed. For the transverse-momentum distribution of the hardest
lepton in \wmza{} for example (Fig.~\ref{fig:ptl1}), we obtain K-factors
between 1.10 and 1.33 over the plotted momentum range.  The large
differential K-factors are therefore caused by the emission of the
additional jet, where the leptonic system recoils against the jet.  The
integrated K-factors are also reduced, namely to $1.23$ for \wpza{} and
$1.29$ for \wmza{} production. 

In Fig.~\ref{fig:RAl1}, we show the separation in the
rapidity--azimuthal-angle plane ($R$ separation) between the photon and
the lepton with the largest transverse momentum. Again, we observe a
significant dependence of the K-factor on the value of the $R$
separation, varying between $1.4$ and $2.55$. Once we include the jet
veto, this dependence is again much smaller.

Therefore, a simple approximation of rescaling the leading-order cross
section with the integrated K-factor does not hold. It is necessary to
perform a full NLO calculation to reproduce the correct shape of the
distributions.

%
%
\section{Conclusions}
\label{sec:concl}
We have calculated the NLO QCD corrections to the processes 
$pp, p\bar{p} \rightarrow W^\pm Z \gamma + X$ including full leptonic decays
of the $W$ and $Z$ boson. With three leptons, a photon and missing transverse
energy as signature, it is an important background for searches
for new physics, in particular supersymmetry. Additionally, it can serve
as a signal process for measuring the quartic gauge couplings
$WWZ\gamma$ and $WW\gamma\gamma$ at the LHC.

We find that the corrections yield a sizable increase of the cross
section with respect to the leading-order result, with integrated
K-factors typically around $1.9$. The LO scale variation strongly
underestimates these contributions with a variation below the $10\%$
level. Varying factorization and renormalization scale $\mu_F = \mu_R =
\mu$ by a factor 2 around the central value $\mu_0 = m_{W Z \gamma}$, we
obtain a remaining scale dependence at NLO of about $7\%$. 

The NLO corrections also show a significant dependence on the observable
and on different phase-space regions. Therefore, it is important to have
a dedicated fully-exclusive NLO parton-level Monte Carlo code available
for \wza{} production. This process will be included into a future
version of the \vbfnlo{} program package.

%
%
\section*{Acknowledgments}
We would like to thank Vera Hankele for helpful
discussions. This research was supported in part by the Deutsche
Forschungsgemeinschaft via the Sonderforschungsbereich/Transregio
SFB/TR-9 ``Computational Particle Physics'' and the Initiative and
Networking Fund of the Helmholtz Association, contract HA-101 
(``Physics at the Terascale''). F.C.~acknowledges partial
support by European FEDER and Spanish MICINN under grant FPA2008-02878.
The Feynman diagrams in this paper were drawn using Jaxodraw~\cite{axo}. 

%
%

\end{document}